\documentclass[12pt]{article}
\usepackage{epsf}
\usepackage{amsmath}
\usepackage{graphics}

\renewcommand{\thefootnote}{\#\arabic{footnote}}

\begin{document}

\setcounter{footnote}{0}
\begin{titlepage}

\begin{center}

\hfill hep-ph/0608115\\
\hfill August 2006\\

\vskip .5in

{\Large \bf
Effects of Cosmic Strings on Free Streaming
}

\vskip .45in

{\large
Tomo Takahashi$^1$ and Masahide Yamaguchi$^2$
}

\vskip .45in

{\em
$^1$Department of Physics, Saga University, Saga 840-8502, Japan \\
$^2$Department of Physics and Mathematics, \\
Aoyama Gakuin University, Sagamihara 229-8558, Japan 
}

\end{center}

\vskip .4in

\begin{abstract}

  We study the effect of free streaming in a universe with cosmic
  strings with time-varying tension as well as with constant
  tension. Although current cosmological observations suggest that
  fluctuation seeded by cosmic strings cannot be the primary source of
  cosmic density fluctuation, some contributions from them are still
  allowed. Since cosmic strings actively produce isocurvature
  fluctuation, the damping of small scale structure via free streaming
  by dark matter particles with large velocity dispersion at the epoch
  of radiation-matter equality is less efficient than that in models
  with conventional adiabatic fluctuation. We discuss its implications
  to the constraints on the properties of particles such as massive
  neutrinos and warm dark matter.

\end{abstract}

{\small PACS numbers; 98.80.Cq}

\end{titlepage}

\renewcommand{\thepage}{\arabic{page}}
\setcounter{page}{1}
\renewcommand{\thefootnote}{\#\arabic{footnote}}

\section{Introduction}

Cosmological observations such as cosmic microwave background (CMB),
large scale structure and so on become now very precise and can
constrain various cosmological parameters with unprecedented accuracy.
In addition to the determination of cosmological parameters such as
energy densities of baryons, dark matter, and dark energy, Hubble
parameter, the scalar spectral index, reionization optical depth and
so on, cosmological observations can also constrain unknowns in
particle physics.  For example, neutrino masses can be severely
constrained by cosmology \cite{Spergel:2006hy,Fukugita:2006rm}. It is
well known that massive neutrinos can erase small scale
inhomogeneities via free streaming since massive neutrinos can have
large velocity dispersion at the time of radiation-matter
equality. Due to this effect, the matter power spectrum exhibits the
suppression at small scales, which can be compared to observations of
large scale structure to give the constraint on neutrino
masses\footnote{
  The mass of neutrino can also affect the CMB power spectrum through
  the modification of the structure of the acoustic peaks, which can
  give a severe constraint on them
  \cite{Fukugita:2006rm,Ichikawa:2004zi}.  }.

Another such example is warm dark matter (WDM) particles. WDM
particles can also have large velocity dispersion at the epoch of
radiation-matter equality like massive neutrinos. WDM scenarios have
been extensively studied in the literature in connection with particle
physics and astrophysics.  As for astrophysical aspects, WDM has been
discussed, in particular, as a solution of the problem of small scale
structure such as the missing satellite problem and the cusp problem
\cite{Colin:2000dn,Bode:2000gq}. From the viewpoint of particle
physics, there exist well-motivated candidates for WDM such as a light
gravitino \cite{Pagels:1981ke,Bond:1982uy}, sterile neutrinos
\cite{Dodelson:1993je} and so on. The properties of WDM such as its
mass can be constrained by cosmological observations as the same
manner as the case with massive neutrinos since WDM particles also
erase the small scale fluctuation via the free streaming effect. It
should also be mentioned that superweakly interacting massive
particles (superWIMPs) \cite{Feng:2003xh} can also erase the small
scale structure as WDM particles, thus this kind of model can also be
constrained by cosmological observations by studying the damping of
matter power spectrum.  Therefore, probing the damping of small scale
structure can be an important test for particle physics.

Constraints on above mentioned particles such as massive neutrinos and
candidates for WDM have been well studied in the framework where cosmic
density fluctuation is seeded by conventional adiabatic primordial
fluctuation which is motivated by inflation driven by a scalar
field. However, cosmic density fluctuation can also be produced by
cosmic strings. Since fluctuation seeded by cosmic strings is an
incoherent actively generated isocurvature one, this kind of fluctuation
cannot produce observed structure of the acoustic peaks in the CMB power
spectrum but gives rise to fairly broad acoustic peaks.  Thus current
observations suggest that cosmic strings cannot be the primary source of
density fluctuation today. However, subdominant contribution from
fluctuation seeded by cosmic strings is still allowed
\cite{Wyman:2005tu}. Furthermore cosmological scenarios with cosmic
strings have been revived for recent years since there have been
discussed that cosmic strings can be produced in a wide class of string
theory models. In particular, cosmic strings can be formed at the end
of brane inflation \cite{Jones:2002cv,Sarangi:2002yt,Jones:2003da}.

In light of these considerations, it is interesting to study the
effect of free streaming in a scenario with cosmic
strings. Importantly, since cosmic strings produce fluctuation
actively, the erasure of small scale inhomogeneities via free
streaming can be avoided to some extent, which results in delayed
damping of small scale power
\cite{Brandenberger:1987er,Brandenberger:1987kf,Bertschinger:1987,Albrecht:1992sb}.
Hence one may consider that some contribution from fluctuation seeded
by cosmic strings can relax the constrains on WDM or neutrino masses
since the constraint on the masses mainly come from the effect of free
streaming. This is the issue which we are going to consider in this
paper.

In fact, the authors of Ref.~\cite{Brandenberger:2004kc} have discussed
the possibilities of relaxing the constraint on neutrino masses in
models with cosmic strings with constant tension. There it was shown by
a simple analytic estimate that even with the addition of fluctuation
from cosmic strings the constraint on neutrino masses cannot be
relaxed. In this paper, first of all, we reconfirm this result with more
quantitative analysis by numerically calculating matter power spectrum
and CMB anisotropy produced by cosmic strings. We also do similar
calculations for WDM and judge whether the constraints on WDM can be
relaxed or not.

Furthermore, recently a new class of cosmic strings has been
considered, which has time-varying tension. When a scalar field
constituting a string couples to another oscillating field or a string
is realized in a brane configuration, cosmic strings with time-varying
tension naturally appear. Cosmological evolution of such a class of
cosmic strings has been investigated in
Refs.~\cite{Yamaguchi:2005gp,Ichikawa:2006rw} where it was shown that,
after some relaxation time, it goes into the scaling regime like
strings with constant tension \cite{ls,gs} and global monopoles
\cite{gm}. In the scaling regime, the typical length of the cosmic
string network grows with the horizon scale. Then, in case that string
tension is constant, the ratio of the energy density of infinite
strings to that of the background universe is also constant, which
generates scale invariant density fluctuations. On the other hand,
when the string tension is time-varying, the ratio of the energy
density of infinite strings to that of the background universe is not
necessarily constant due to the time dependence of the tension. For
example, when the tension has the time dependence as $G\mu \propto
\tau^{n}$ with $\tau$ being the conformal time, fluctuations at small
scales are enhanced but those at large scales are suppressed for
negative values of $n$. Thus the above discussion on the effects of
cosmic strings on free streaming may be modified when we consider the
case with time-varying tension.

The purpose of this paper is to study to what extent fluctuation from
cosmic string with time-varying tension as well as with constant
tension can affect to avoid the erasure of small scale inhomogeneities
via free streaming by massive neutrinos and WDM. We also discuss the
implications of the above mentioned phenomenon to the constraints on
the masses of these particles.

\section{Free streaming effect in a universe with cosmic strings }

Although fluctuation from cosmic strings cannot be the primary source
of cosmic density fluctuation today, some contributions from them are
still allowed. As mentioned in the introduction, the damping of small
scale fluctuation by free streaming can be avoided to some extent when
fluctuation is seeded by cosmic strings.  Thus, even in models with
massive neutrinos or WDM, the erasure of small scale fluctuation can be
less efficient by adding some fluctuation produced from cosmic
strings, which may have much implications to the constraints on the
masses of neutrinos and WDM.

First we discuss this issue for the case with massive neutrinos. In
fact, the discussion on the alleviation of the constraint on neutrino
masses has already been made in Ref.~\cite{Brandenberger:2004kc}, where
it was shown that even with the addition of fluctuation from cosmic
strings the constraint on neutrino masses cannot be relaxed. This is
explained as follows. In order that fluctuation from cosmic strings can
affect the matter power spectrum at small scales where the free
streaming effect erases the inhomogeneities, the amplitude of the
fluctuation should be as large as that of the conventional adiabatic
one.  In this case, however, the fluctuation amplitude becomes too large
at larger scales where the current CMB measurements are relevant.
Notice that the fluctuation from cosmic string becomes an isocurvature
one which gives the Sachs-Wolfe (SW) effect on the CMB temperature
anisotropies as
\begin{equation}
\displaystyle\frac{\Delta T}{T} \Bigr|_{\rm SW}  =  2 \Phi,
\end{equation}
where $\Phi$ is the gravitational potential which appears in a metric
perturbation in the conformal Newtonian gauge. On the other hand, the SW
effect in the conventional adiabatic case can be written as
\begin{equation}
\displaystyle\frac{\Delta T}{T} \Bigr|_{\rm SW}  =  \frac{1}{3} \Phi.
\end{equation}
Thus for the same magnitude of $\Phi$, the isocurvature fluctuation
can give about 6 times larger CMB temperature fluctuation on large
scales than that from the adiabatic one.  In other words, fixing the
amplitude to give the right size to fit the CMB data, matter power
spectrum for isocurvature fluctuation should be about 36 times smaller
than that for adiabatic fluctuation. Thus, it is impossible to relax
the constraint on neutrino masses using the isocurvature fluctuation
seeded by cosmic strings without affecting the CMB constraint.

\begin{figure}[t]
\begin{center}
\scalebox{0.8}{\includegraphics{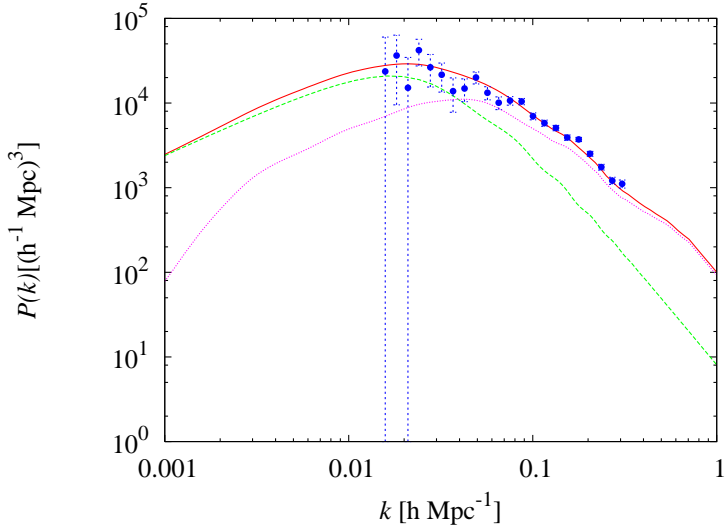}}
\caption{Matter power spectrum in models with massive neutrinos ($\sum
  m_{\nu} = 5$ eV) for the cases with conventional adiabatic
  fluctuation (green dashed line), cosmic strings with constant
  tension $G\mu = 5.2 \times 10^{-6}$ (purple dotted line) and
  the total matter spectrum (red solid line). For reference, we also
  plot the data from SDSS\cite{Tegmark:2003uf}.}
\label{fig:Pk_mnu5}
\end{center}
\end{figure}

\begin{figure}[htbp]
\begin{center}
\scalebox{0.8}{\includegraphics{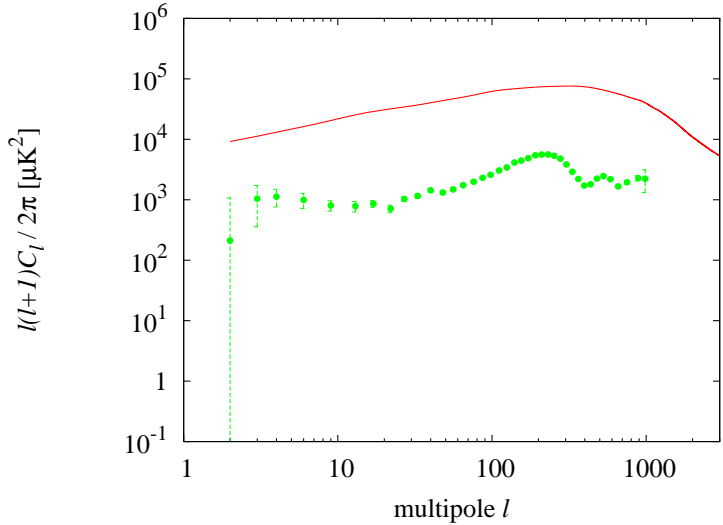}}
\caption{The CMB TT power spectrum in models with massive neutrinos
  ($\sum m_{\nu} = 5$ eV) for the case with cosmic strings with
  constant tension $G\mu = 5.2 \times 10^{-6}$ (solid red
  line). The data from WMAP3 are also plotted \cite{Spergel:2006hy}.
}
\label{fig:Cl_mnu5}
\end{center}
\end{figure}

Here we demonstrate this quantitatively by calculating the matter and
CMB power spectra in models with massive neutrinos and cosmic stings
using the CMBACT code \cite{Pogosian:1999np}.  In
Fig.~\ref{fig:Pk_mnu5}, we plot the matter power spectra $P(k)$ for
the case with the conventional adiabatic fluctuation $P(k)_{\rm
  (adiabatic)}$ (green dashed line), cosmic strings with constant
tension $P(k)_{\rm (string)}$(purple dotted line) and the total power
$P(k)_{\rm ( adiabatic)} + P(k)_{\rm (string)}$ (red solid line)
assuming neutrino masses as $\sum m_{\nu} = 5$ eV\footnote{
  Neutrino masses can be constrained from CMB data alone through the
  modification of the structure of acoustic peaks and the current
  bound is $\sum m_{\nu} < 2.0$ eV at 95\%
  C.L. \cite{Fukugita:2006rm}. However this bound may not hold true in
  our case with not only adiabatic fluctuation but also isocurvature
  fluctuation generated by cosmic strings. Thus, for an illustration,
  we assumed this value for neutrino masses.  }.
For reference, we also plot the data from SDSS \cite{Tegmark:2003uf}.
The other cosmological parameters are assumed as $\Omega_{\rm m}h^2=
0.139, \Omega_bh^2= 0.021, h = 0.66, n_s =1$ and $\tau_{\rm reion} =
0.091$ where $\Omega_{m,b} $ are the present energy densities
normalized by the critical density for matter and baryon, $h$ is the
Hubble parameter, $n_s$ is the scalar spectral index and $\tau_{\rm
  reion}$ is the reionization optical depth. A flat universe is
assumed. The energy density of massive neutrinos are added by reducing
that of dark energy.  We adopt these values unless otherwise stated in
this paper.  As seen from the figure, although the matter power
spectrum for adiabatic fluctuation is damped on small scales in this
case, we can compensate the damping by adding fluctuation from comics
strings with constant tension $G\mu = 5.2 \times 10^{-6}$.  However
the CMB power spectrum generated by the cosmic strings with the size
of the tension has too much power and contradicts with WMAP
observations, as shown in Fig. \ref{fig:Cl_mnu5}.

In the discussion above, we considered cosmic strings with constant
tension. Recently, however, cosmic strings with time-varying tension
have been studied in Refs.~\cite{Yamaguchi:2005gp,Ichikawa:2006rw}. In
particular, in Ref.~\cite{Ichikawa:2006rw}, the CMB and matter power
spectra have been discussed for such models. It was explicitly shown
that for the cases where the string tension decreases with time as
$G\mu \propto \tau^{n}$ or $\propto a^{n}$ with $a$ being the scale
factor and $n$ being negative power, the fluctuation on large scales
is significantly suppressed and that on small scales is
enhanced. Hence we can naively expect that the above argument on the
alleviation of the constraint on neutrino masses can be modified for
such cosmic strings. We studied this issue by calculating the CMB and
matter power spectrum using the modified version of CMBACT code where
we have introduced the time dependence of the tension as $G\mu \propto
\tau^{n}$ or $\propto a^{n}$.

\begin{figure}[htbp]
\begin{center}
\scalebox{0.8}{\includegraphics{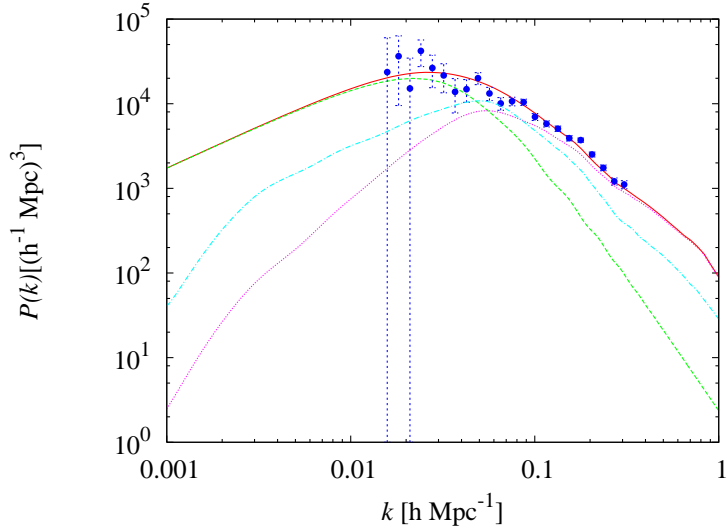}}
\caption{Matter power spectrum in models with massive neutrinos ($\sum
  m_{\nu} = 10$ eV) for the cases with conventional adiabatic
  fluctuation (green dashed line), cosmic strings with time-varying
  tension $G\mu \propto \tau^{-0.4}$ (purple dotted line) and the
  total power spectrum from the adiabatic fluctuation plus cosmic
  strings with time-varying tension (red solid line). For comparison,
  the case with constant tension $G\mu = 5.1 \times 10^{-6}$ is also
plotted (cyan dash-dotted line).}
\label{fig:Pk_mnu10_ct}
\end{center}
\end{figure}

\begin{figure}[htbp]
\begin{center}
\scalebox{0.8}{\includegraphics{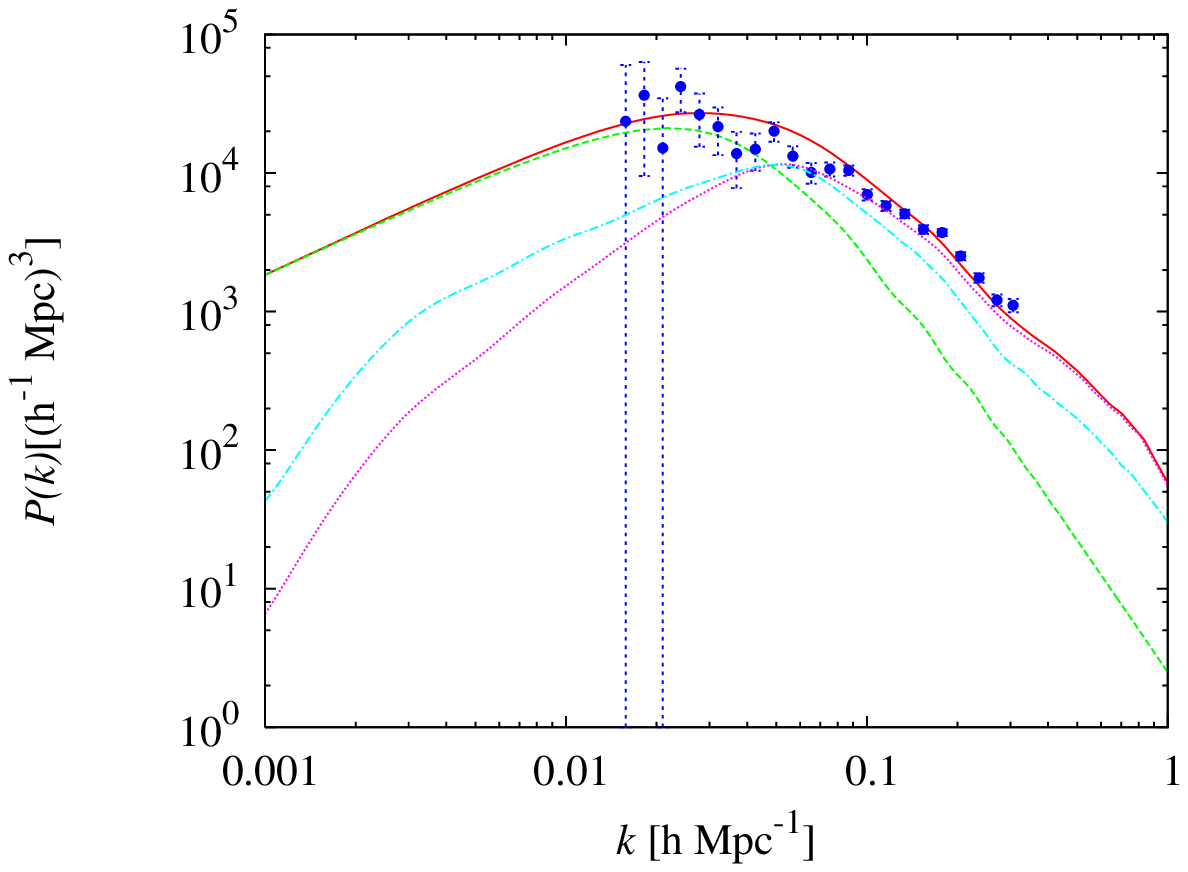}}
\caption{Matter power spectrum in models with massive neutrinos ($\sum
  m_{\nu} = 10$ eV) for the cases with conventional adiabatic
  fluctuation (green dashed line), cosmic strings with time-varying
  tension $G\mu \propto a^{-0.2}$ (purple dotted line) and the total
  power spectrum from the adiabatic fluctuation plus cosmic strings
  with time-varying tension (red solid line). For comparison, the case
  with constant tension $G\mu = 5.2  \times 10^{-6}$ is also plotted
  (cyan dash-dotted line).}
\label{fig:Pk_mnu10_sf}
\end{center}
\end{figure}

\begin{figure}[htbp]
\begin{center}
\scalebox{0.8}{\includegraphics{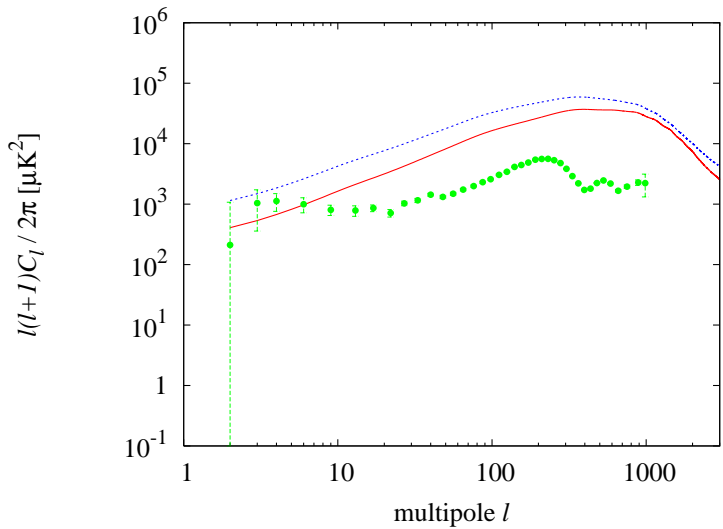}}
\caption{The CMB TT power spectrum in models with massive neutrinos
  ($\sum m_{\nu} = 10$ eV) for the case with cosmic strings with
  time-varying tension $G\mu \propto \tau^{-0.4}$ (solid red line) and
  $G\mu \propto a^{-0.2}$ (blue dashed line). The data from WMAP3 are
  also plotted \cite{Spergel:2006hy}.}
\label{fig:Cl_mnu10}
\end{center}
\end{figure}

As an example, in Fig.~\ref{fig:Pk_mnu10_ct}, we plot the matter power
spectrum in models with massive neutrinos ($\sum m_{\nu} = 10$ eV) for
the cases with with conventional adiabatic fluctuation (green dashed
line), cosmic strings with time-varying tension $G\mu \propto
\tau^{-0.4}$ (purple dotted line) and the total power spectrum from
these fluctuations (red solid line).  For comparison, we also plot the
case with cosmic strings with constant tension (cyan dash-dotted
line).  Another case for the time dependence of the string tension
with $G\mu \propto a^{-0.2}$ is also shown in
Fig.~\ref{fig:Pk_mnu10_sf}.  As seen from the figure, fluctuation from
cosmic strings with the time dependence $G\mu \propto \tau^{-0.4}$ and
$G\mu \propto a^{-0.2}$ can cancel the damping caused by free
streaming of massive neutrinos. However, even in these cases, since
adiabatic fluctuation with massive neutrinos damps the matter power
spectrum on smaller scales than around $k \sim 0.02 {\rm Mpc}^{-1}$,
we have to add the contribution from cosmic strings to compensate the
damping around this scale.  When we have non-negligible amplitude from
cosmic strings around this scale, it generates too much power to fit
the CMB data even though the fluctuation on large scales is reduced
due to the time dependence of the sting tension.  To show this, in
Fig. \ref{fig:Cl_mnu10}, we plot the CMB power spectrum generated by
cosmic string with time varying tension $G\mu \propto \tau^{-0.4}$ and
$a^{-0.2}$ with the same normalization (string tension) as that in
Fig.~\ref{fig:Pk_mnu10_ct}.  As seen from the figure, we cannot
compensate the erasure of fluctuation on such scales without
contributing to the CMB power spectrum significantly, which is
obviously inconsistent with current observations, even if we introduce
the cosmic strings with time-varying tension.  We note that, depending
on the neutrino masses, we can compensate the damping of small scale
power via free streaming by choosing the time dependence of the string
tension appropriately.  However, our conclusion remains unchanged as
long as adiabatic fluctuation with massive neutrinos damps the matter
power spectrum on small scales which correspond to $\ell \sim
\mathcal{O}(100)$.


\begin{figure}[htbp]
\begin{center}
\scalebox{0.8}{\includegraphics{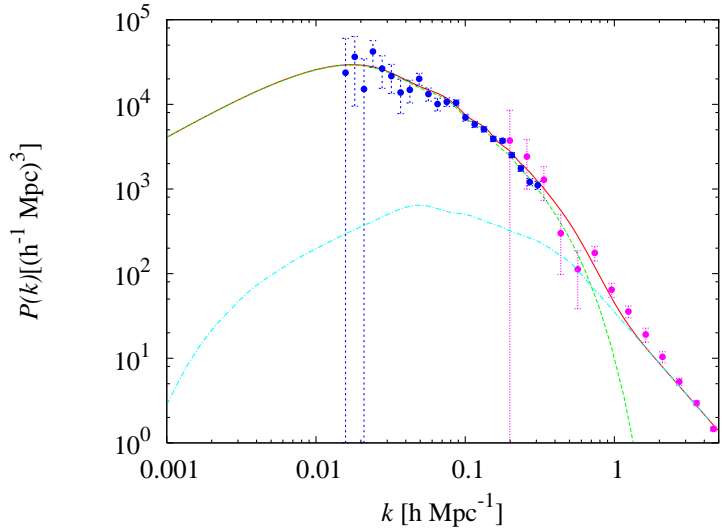}}
\caption{Matter power spectrum in models with WDM with the mass
  $m_{\rm WDM} = 103$ eV.  Here we plot $P(k)$ from the conventional
  adiabatic fluctuation (green dashed line), cosmic strings with
  constant tension $G\mu = 7.8 \times 10^{-7}$ (cyan dot-dashed line)
  and the total of them (solid red line).  For reference, the data
  from SDSS \cite{Tegmark:2003uf} (blue points) and Lyman alpha
  \cite{Croft:2000hs} (purple points) which are
  rescaled to $z=0$ are also plotted.}
\label{fig:Pk_WDM100}
\end{center}
\end{figure}

\begin{figure}[htbp]
\begin{center}
\scalebox{0.8}{\includegraphics{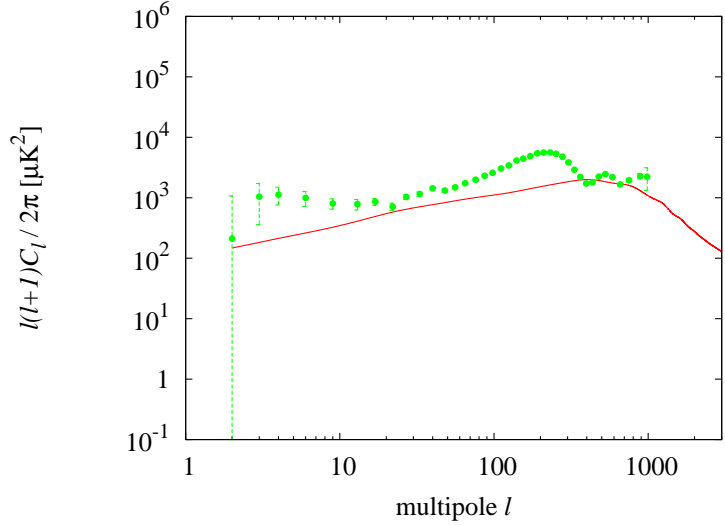}}
\caption{The CMB TT power spectrum in WDM models with the mass $m_{\rm
    WDM} = 103$ eV for the case with cosmic strings with constant
  tension $G\mu = 7.8 \times 10^{-7}$ (solid red line). The data
  from WMAP3 are also plotted \cite{Spergel:2006hy}.}
\label{fig:Cl_WDM100}
\end{center}
\end{figure}

Next we discuss the effects of cosmic strings on free streaming for
the case with WDM. For WDM, since the scales of masses are
significantly larger than those of massive neutrinos considered above,
the damping scale can be much smaller. Thus in this case, we may have
a chance to avoid the erasure of small scale inhomogeneities by free
streaming without affecting the CMB power spectrum on large scales by
adding isocurvature fluctuation from cosmic strings. In
Fig.~\ref{fig:Pk_WDM100}, we plot the matter power spectrum for the
cases with conventional adiabatic fluctuation (green dashed line),
cosmic strings with constant tension $G\mu = 7.8 \times 10^{-7}$ (cyan
dot-dashed line) and the total power spectrum (red solid line). For
reference, we have also plotted the data from SDSS
\cite{Tegmark:2003uf} and Lyman alpha \cite{Croft:2000hs} which are
rescaled to the present time $z=0$. Here we assumed the pure WDM
model.  The mass of WDM is taken to be $m_{\rm WDM} = 103$ eV, which
corresponds to a thermally decoupled relic with the relativistic
degrees of freedom at the time of decoupling being $g_*(T_D) = 100$.
For the case with adiabatic fluctuation, WDM particles erase the
structure on smaller scales compared to the case with massive
neutrinos as seen from the figure.  With the parameter above, the
damping scale is $k \sim 1 {\rm Mpc}^{-1}$. On the other hand, for the
case with cosmic strings, such a damping becomes moderate.  As shown
in Fig. \ref{fig:Pk_WDM100}, the small scale damping for adiabatic
fluctuation can be canceled by adding that from cosmic strings with
constant tension. Furthermore, the CMB power spectrum generated by the
above cosmic strings can satisfy WMAP constraints, as shown in
Fig. \ref{fig:Cl_WDM100}. Thus the constraint on WDM can be relaxed in
a scenario with comics strings even with constant tension.

\begin{figure}[htbp]
\begin{center}
\scalebox{0.8}{\includegraphics{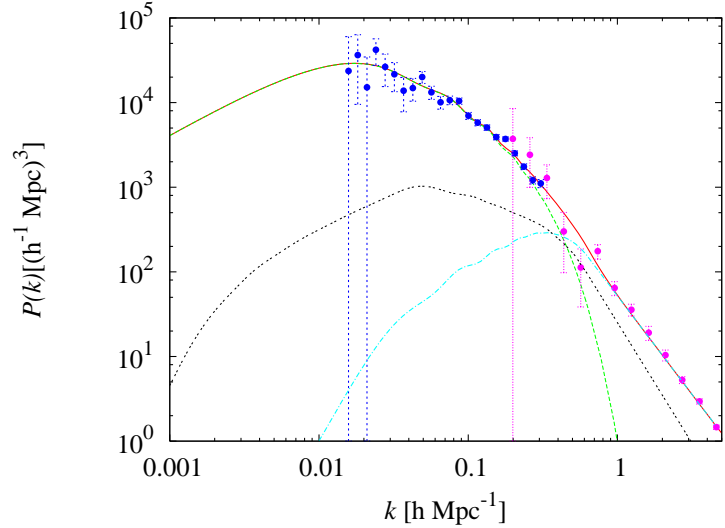}}
\caption{Matter power spectrum in models with WDM with the mass
  $m_{\rm WDM} = 72$ eV.  Here we plot $P(k)$ from the conventional
  adiabatic fluctuation (green dashed line), cosmic strings with
  time-varying tension $G\mu \propto \tau^{-1}$ (cyan dot-dashed
  line), and the total matter spectrum from the adiabatic fluctuation
  plus cosmic string with time-varying tension (solid red line).  The
  case with constant tension $G\mu = 9.8 \times 10^{-7}$ is also
  plotted for comparison (black dash-dashed line).  For reference, the
  data from SDSS \cite{Tegmark:2003uf} (blue points) and Lyman alpha
  \cite{Croft:2000hs} (purple points) which are rescaled to $z=0$ are
  also plotted.  }
\label{fig:Pk_WDM70}
\end{center}
\end{figure}

\begin{figure}[htbp]
\begin{center}
\scalebox{0.8}{\includegraphics{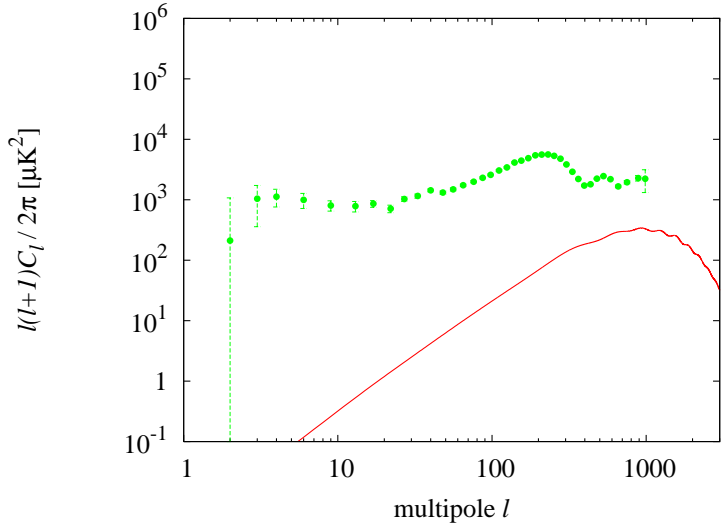}}
\caption{The CMB TT power spectrum in WDM models with the mass $m_{\rm
    WDM} = 72$ eV for the case with cosmic strings cosmic strings with
  time-varying tension $G\mu \propto \tau^{-1}$ (solid red line). The
  data from WMAP3 are also plotted \cite{Spergel:2006hy}.}
\label{fig:Cl_WDM70}
\end{center}
\end{figure}

In addition, we also show matter power spectrum for the case with
cosmic strings with time-varying tension $G\mu \propto \tau^{-1}$,
conventional adiabatic fluctuation and the total spectrum of them in
Fig~\ref{fig:Pk_WDM70}.  Here we assume the mass of WDM particles as
$m_{\rm WDM} \sim 72$ eV.  As seen from the figure, we can cancel the
damping on small scales in the spectrum for adiabatic case by adding
the contribution from cosmic strings with the time-varying tension
$n=-1$. Importantly, in this case, the contribution from the cosmic
string to CMB power spectrum up to $\ell \sim \mathcal{O}(1000)$ is
negligible since large scale fluctuation by the cosmic strings is
significantly suppressed due to the time dependence of the string
tension (Fig. \ref{fig:Cl_WDM70}).  Hence we can have possibilities of
alleviating the constraint on the masses of WDM by adding some
fluctuation from comics strings with time-varying tension in this case
too.

\section{Conclusion and Discussion}

We studied the erasure of small scale structure via free streaming in
models with cosmic strings with time-varying tension as well as with
constant tension. Because cosmic strings actively produce incoherent
isocurvature fluctuation, free streaming effect is less efficient than
that in models with the conventional adiabatic ones because of its
nature of the fluctuation. Since the damping of small scale power by
free streaming is an important probe to constrain the property of
particles with large velocity dispersion at the epoch of
radiation-matter equality such as massive neutrinos and WDM, above
mentioned effects by cosmic strings can have much implications to the
constraints on such particles.

We first studied this issue for massive neutrinos. For the case with
constant string tension, it has been already discussed some time ago
that, even if we add some contribution from isocurvature fluctuation
from cosmic strings, we cannot relax the constraint on neutrino
masses. In this paper, we explicitly calculate matter power spectrum
and CMB anisotropy generated by strings and reach the same
conclusion. Furthermore, we have studied this issue for the case with
cosmic strings with time-varying tension. We have shown that even if
we introduce cosmic strings with time-varying tension, we cannot
alleviate the constraint on the mass without conflicting with the
constraint from current CMB observations.  This is mainly because
adiabatic fluctuation with massive neutrinos damps the matter power
spectrum on small scales corresponding to multipoles 
$\ell \sim \mathcal{O}(100)$ in CMB power spectrum.

We have also discussed the possibilities of relaxing constraints on
the masses of WDM by adding fluctuation produced by cosmic strings
with time-varying tension and those with constant tension. We
explicitly showed that the damping of the matter power spectrum on
small scales from adiabatic primordial fluctuation by free streaming
caused by WDM can be canceled by adding fluctuation seeded by cosmic
strings for both cases, namely with constant and time-varying tension
by choosing appropriate parameters.  Importantly we can cancel the
damping without conflicting the CMB constraints for both cases, which
is not the case for massive neutrinos.  Although much more detailed
study is needed to state qualitatively to what extent the constraint
on WDM particles can be relaxed, in this paper, we pointed out that
some contributions from cosmic strings to small scale power can affect
the constraints without conflicting with the CMB constraint.  Since
the time dependence of the string tension is highly model-dependent
and can be more complicated than that adopted here, various
possibilities can arise and have much more implications to the
constraints given from observations probing the damping of small scale
fluctuation via free streaming.

\section*{Acknowledgments}

We would like to acknowledge the use of CMBACT code developed and made
publicly available by L. Pogosian and T. Vachaspati. M.Y. is supported
in part by the project of the Research Institute of Aoyama Gakuin
University and by the JSPS Grant-in-Aid for Scientific Research No.\
18740157.

\end{document}